\documentclass[11pt]{article}
\usepackage{moriond,epsfig,graphicx,amssymb,psfig}

\def \aj {AJ}
\def \mnras {MNRAS}
\def \apj {ApJ}
\def \apjs {ApJS}
\def \apjl {ApJL}
\def \aap {A\&A}

\def\be{\begin{equation}}
\def\ee{\end{equation}}
\def\bea{\begin{eqnarray}}
\def\eea{\end{eqnarray}}

\def \zabs {z_{\rm abs}}

\def \caii {Ca~{\sc ii}}

\def \znii {Zn~{\sc ii}}

\def \mgii {Mg~{\sc ii}}

\def \hi {H~{\sc i}}
\def \hii {H~{\sc ii}}
\def \nhi {$N$(H~{\sc i})}

\def \EBV {\textit{E(B$-$V)}}
\def \oii {[O~{\sc ii}]}
\def \kms {\rm km\,s^{-1}}
\def \sfrd {\dot{\rho}^*}
\def \ergscm {\rm ergs\,cm^{-2}\,s^{-1}}
\def \Msolyr {\rm M_\odot yr^{-1}}
\def \perkpcsq {\rm kpc^{-2}}

\def\gsim{\mathrel{\lower0.6ex\hbox{$\buildrel {\textstyle >}
 \over {\scriptstyle \sim}$}}}
\def\lsim{\mathrel{\lower0.6ex\hbox{$\buildrel {\textstyle <}
 \over {\scriptstyle \sim}$}}}
 
\begin{document}
\vspace*{4cm}
\title{\caii\ and DLA ABSORPTION LINE SYSTEMS: DUST, METALS AND 
STAR FORMATION AT $0.4<z<1.3$}

\author{VIVIENNE WILD$^1$, PAUL HEWETT$^2$, MAX PETTINI$^2$}

\address{$^1$Max Planck Institute for Astrophysics, Karl-Schwarzschild
  Str. 1, 85748 Garching, Germany \\
$^2$Institute of Astronomy, University of Cambridge, Madingley Road,
Cambridge CB3 0HA, UK\\}

\maketitle

\abstracts{Absorption line studies of galaxies along the line-of-sight
to distant quasars allow a direct observational link between the
properties of the extended gaseous disk/halo and of the star forming
region of galaxies.  In these proceedings we review recent work on
\caii\ absorbers detected in the SDSS at $0.4<z<1.3$ which, because of
their dust content and chemical properties, may lie spatially closer
to the central host galaxy than most DLAs.  We present direct evidence for the
presence of star formation, through observation of the
\oii\,$\lambda\lambda$3727,3730 emission line, in both \caii \
absorbers and \mgii-selected Damped-Lyman-$\alpha$ (DLA) systems.  The
measured SFR from light falling within the SDSS fibre apertures
(corresponding to physical radii of $6-9 h^{-1}\,$kpc) is
$0.11-0.48\,\Msolyr$ for the \caii-absorbers and $0.11-0.14\,\Msolyr$
for the \mgii-selected DLAs.  The contribution of both \caii \
absorbers and DLAs to the total observed star formation rate density,
$\sfrd$, in the redshift range $0.4 < z < 1.3$, is small, $<10\%$.
Our result contrasts with recent conclusions, based on the Schmidt
law, that DLA absorbers can account for the majority of the total
observed $\sfrd$ in the same redshift range.}

\section{Introduction}

% Absorption line systems
The strong intervening hydrogen and metal absorption lines observed in
quasar spectra are generally thought to be caused by gas associated
with galaxies along the line-of-sight. Damped Lyman-$\alpha$ (DLA)
systems are the most extreme subset of these absorbers, defined
through their high column densities, \nhi$>2 \times 10^{20}\,{\rm cm}^{-2}$,
of neutral hydrogen
gas.

% Why are they interesting? bridging the gap
% Relevance for this conference

In principle, galaxies detected through the absorption of background
quasar light by their interstellar gas provide a unique view of the
chemical evolution of galaxies and their outer gaseous halos over the
majority of the age of the universe, unaffected by the luminosity bias
suffered by traditional emission-selected galaxy samples. They present
the opportunity to accurately probe metallicity, dust attenuation,
disk dynamics to large radii and gas cross-section weighted star
formation.  Indeed, accurate chemical abundances of these absorbers
are already being used to provide useful comparisons to SPH galaxy
evolution simulations\,\cite{2004MNRAS.348..435N}.

% Problem is...
However, the true nature of absorption systems is unclear: their
persistently low metallicities\,\cite{2005ApJ...618...68K,2003ApJ...595L...9P} and dust
contents\,\cite{2004MNRAS.354L..31M} may suggest they exist within
galaxies with a low degree of chemical evolution, or, probe only the
outer, relatively metal poor, parts of disks. Follow-up deep imaging
campaigns have had mixed success with regard to identification of the
host galaxy responsible for the absorber. At high redshift, interest
has focussed on the DLAs, where only a handful of suitable candidates
have been detected\,\cite{2005MNRAS.358..985W}, generally through
Lyman-$\alpha$ emission. At lower redshift far fewer DLAs are known,
as the Lyman-$\alpha$ line moves into the ultraviolet, only accessible
to (the now inactive) STIS on the Hubble Space Telescope. The few host
galaxies identified appear to be morphologically diverse and span a
wide range of luminosities\,\cite{2005MNRAS.364.1467Z}. A residual
concern is the impact of dust obscuration bias: it is possible that
the most metal rich, and therefore dustiest, absorbers are being
missed from magnitude-limited quasar surveys, although DLAs with such
extreme properties have proved elusive so far\,\cite{CJA05} (see also
contribution by C. Peroux in these proceedings).  The large gap in our
understanding that remains between the absorption- and
emission-selected galaxy populations severely restricts our ability to
gain a more complete understanding of galaxy evolution.

\subsection{\caii\ absorbers}
% Introduce CaII absorbers...

Recently, a different class of absorption line system has received
renewed attention, the \caii\ absorbers \cite{2005MNRAS.361L..30W}.
Although extremely rare (number density about 30\% that of DLAs), a
large sample with $0.4<\zabs<1.3$ has now been identified in the Sloan
Digital Sky Survey (SDSS) DR4 spectroscopic database.  In the local
interstellar medium (ISM) calcium is severely depleted onto dust
grains, and \caii\ is a minor ionisation state with an ionisation
potential below that of hydrogen.  Significant columns of \caii\ are
therefore expected to trace either, i) very high column density DLAs,
ii) high {\it volume} density regions where a degree of self shielding
can take place, or, iii) shocked gas in which dust grain destruction
temporarily releases relatively large quantities of calcium into the
gas phase.  In these proceedings we review the unusual properties of
the absorbers and present new measurements of their star formation
rate (SFR), from detection of \oii\,$\lambda\lambda3727,3730$
in stacked SDSS spectra.  We conclude by discussing some implications of our
results for calculating the star formation rate in \caii \ and DLA
absorbers and their contribution to the total volume averaged SFR
density, $\sfrd$, of the universe at $z\sim 1$.

%%%%%%%%%%%%%%%%%%%%%%%%%%%%%%%%%%%%%%%%%%%%%%%%%%%%%%%%%%%%%%%%%%

\section{Reddening by Dust}

\begin{figure}
\centering
%\psfig{figure=../../CaII_p2/SUBMIT2/ca_dustdr4.ps,height=1.5in}\\
%\vspace*{-0.5cm}
%\psfig{figure=../../CaII_p2/SUBMIT2/ca_dustsplitdr4.ps,height=1.5in}
\psfig{figure=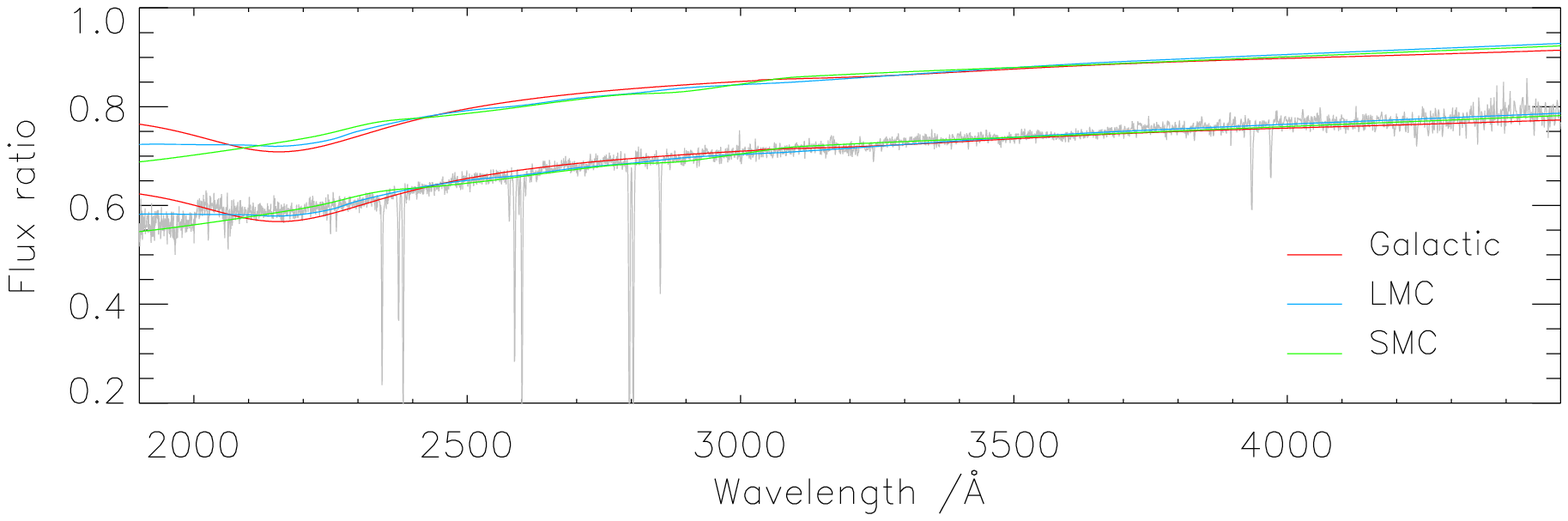,height=1.5in}\\
\vspace*{-0.5cm}
\psfig{figure=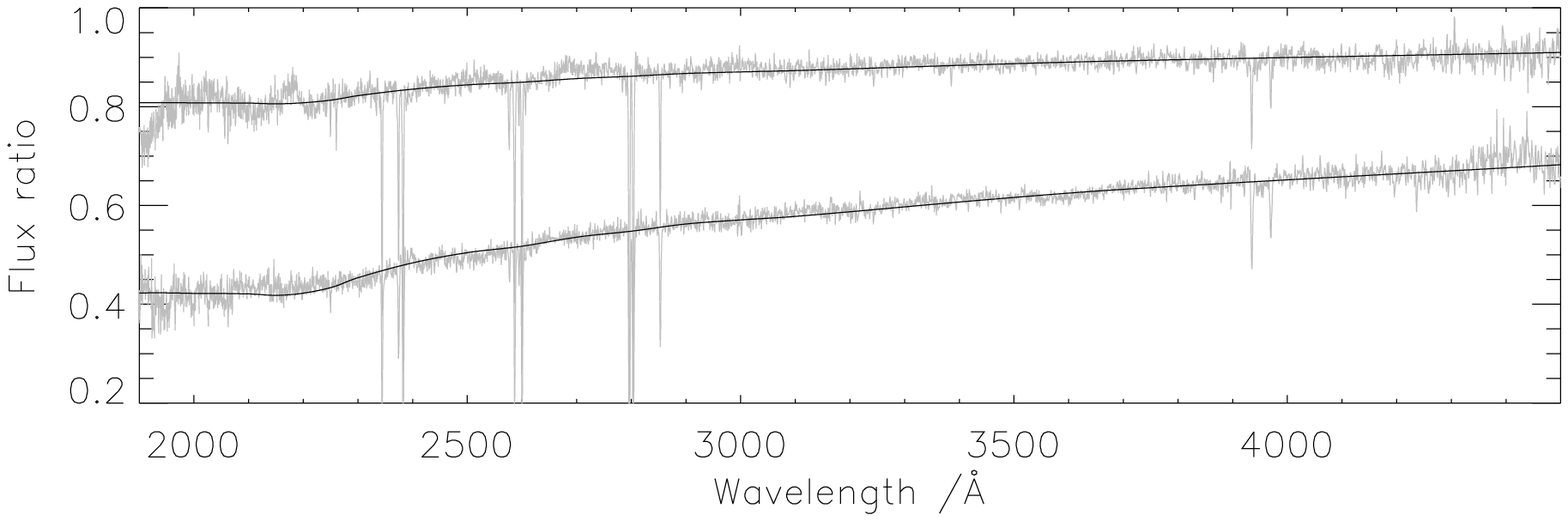,height=1.5in}
\caption{\emph{Top}: Composite spectrum of the 37 quasars with \caii\
absorbers, after division by an `average' quasar SED. Overplotted are
best-fit extinction curves with \EBV$=0.057,0.065,0.066$ for the MW,
LMC and SMC respectively \emph{Bottom}: Same as for the upper panel,
but with the \caii\ sample now split into the `High-' and
`Low-EW' subsamples ($0.3<W_{\lambda3935}<0.68$\,\AA\
and $W_{\lambda3935}>0.68$\,\AA).  The former shows the higher
extinction with \EBV$=0.103$. Data taken from Wild, Hewett \& Pettini
(2006a)$^9$.}
\label{fig:red}
\end{figure}

Dust has three observable effects in galaxies: it selectively depletes
metals in the ISM, reddens their spectral energy distributions, and
causes an overall extinction of the light.  Estimates of reddening
from differences in the average spectral energy distributions (SEDs)
of background quasars must rely on statistical studies, due to the
intrinsic range in the SEDs of quasar spectra.  Due, in part, to the
difficulty in defining suitable samples, there are relatively few
reports of evidence for dust in DLAs via the reddening effect on
background quasar spectra.  However, the large sample of quasars in
the SDSS at similar redshifts to those containing absorption line
systems allows a good estimate of the average quasar spectrum, and
scatter, to be obtained.  Any reddening signal in the spectra of
quasars with an intervening absorber can then be found by comparing
their SEDs with that of the average quasar SED at the appropriate
redshift.

Figure \ref{fig:red} presents the reddening curves resulting from
combining, in the absorber rest frame, 37 \caii\ absorbers with
$0.8<\zabs<1.3$, after each contributing spectrum has been divided by
a suitable high signal-to-noise ratio (SNR) `average' quasar composite
spectrum. The resulting composite is fitted with extinction curves
appropriate to dust in the Milky Way and the Magellanic Clouds (as
indicated) to deduce the values of the colour excess, \EBV.  The
nominally better fit of the LMC dust curve to the high equivalent
width (EW) \caii\ sample ($W_{\lambda 3935}>0.68$\,\AA) suggests the
presence of a weak 2175\AA\ dust feature, although, with the currently
limited statistics, this result is only tentative.

%%%%%%%%%%%%%%%%%%%%%%%%%%%%%%%%%%%%%%%%%%%%%%%%%%%%%%%%%%%%%%%%%%

\vspace{-0.5cm}

\section{Element Depletions}

\begin{figure}
\begin{center}
\psfig{figure=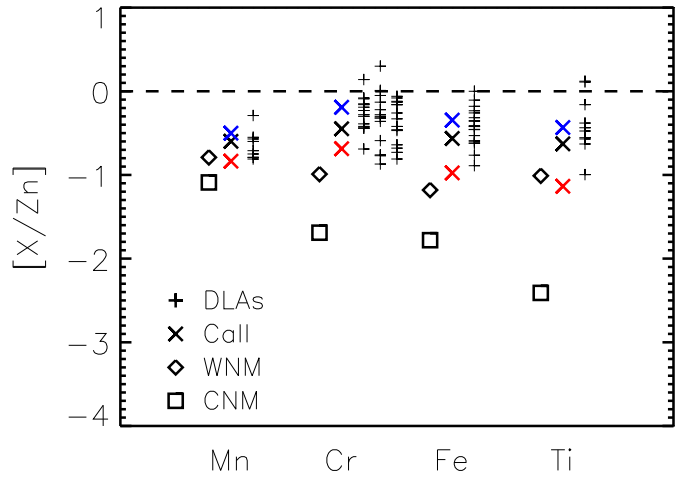,height=3.in}
\end{center}
\caption{The abundances of refractory elements relative to \znii\ in
the warm and neutral ISM of the Milky Way$^{12}$
and DLAs compared to the \caii\ absorbers. The upper, middle, and
lower \caii\ points are for the low-EW, combined, and high-EW samples
respectively of Wild, Hewett \& Pettini (2006a)$^9$, to which the reader
is referred for the sources of the DLA measurements.}
\label{fig:depl} 
\end{figure}

A further diagnostic of the presence of dust in DLAs and metal
absorption line systems comes from relative abundances of elements,
such as Cr to Zn, which are depleted by differing amounts onto dust
grains.  Previous results have indicated that dust depletion is far
less severe than in the Galactic ISM today which has
further strengthened the argument that DLAs are relatively dust free
compared with modern galaxies\,\cite{1997ApJ...486..665P}.

While very weak metal transition lines cannot be seen in individual
SDSS absorption spectra, by combining all the \caii\ absorption
spectra into a single composite, a high enough SNR can be reached to
obtain reliable average measurements.  These average values can be
interpreted readily providing that: (a) the gas in the \caii\
absorbers is predominantly neutral, as is the case in DLAs, and (b)
the distribution of equivalent widths of the lines among the
individual systems is reasonably uniform, as suggested by inspection
of the individual spectra.

Figure \ref{fig:depl} shows ion column densities relative to \znii
\,\footnote{These are quoted in the standard way, relative to solar
abundances\cite{2003ApJ...591.1220L}: ${\rm [X/Zn]} \equiv \log{\rm
[N(X)/N(Zn)]} -\log{\rm [X/Zn]}_\odot$.} \ measured in a composite of
27 \caii\ absorbers with $0.84<\zabs<1.3$, compared to results from
the ISM of the MW and DLAs. \znii\ is used for normalisation as it is
expected to suffer minimal depletion onto dust grains. The depletion
pattern of the \caii\ absorbers is similar to that seen in DLAs, but
the overall level of depletion of the high-EW sample is higher,
approaching values typical of the warm neutral medium of the Milky
Way\,\cite{1999ApJS..124..465W}.

%%%%%%%%%%%%%%%%%%%%%%%%%%%%%%%%%%%%%%%%%%%%%%%%%%%%%%%%%%%%%%%%%%

\section{Star formation from \oii\,$\lambda\lambda$3727,3730 emission}

\begin{figure}
\begin{center}
\psfig{figure=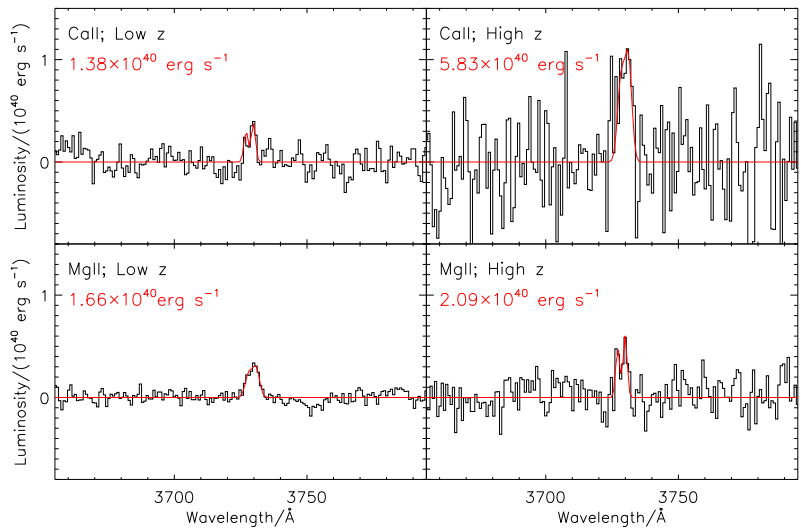,height=3.in}
\end{center}
\caption{SDSS composite spectra of \caii\ absorption line systems
  (top) and \mgii-selected DLA candidates (bottom) showing the region
  of \oii\,$\lambda\lambda$3727,3730 in redshift bins of $0.4 \le
  \zabs < 0.8$ (left) and $0.8 \le \zabs < 1.3$ (right). Flux from
  each contributing spectrum has been converted into luminosity before
  combining. Overplotted (in red lines) are two-component
  Gaussian fits to the emission lines. Measured line luminosities from
  these line fits are given in each plot. Data taken from Wild, Hewett
  \& Pettini (2006b)$^{13}$.}
  \label{fig:sfr}
\end{figure}

A complete picture relating the gas cross-section, ionisation state
and chemical abundances of the ISM of galaxies to their
stellar and nebular emission properties requires knowledge of the SFRs
of absorption-selected galaxies.  However, despite considerable
effort, deep follow-up imaging and spectroscopy have so far led to the
identification of only small numbers of host galaxies of DLAs.  

A significant breakthrough in this field is the recent detection of
\oii\,$\lambda\lambda 3727,3730$ nebular emission associated
with \caii- and \mgii-selected DLAs in stacked spectra
of SDSS QSOs\,\cite{Ca_SFR}. The 1$\sigma$ noise in the SDSS quasar
spectra is $\sim 1 \times 10^{-17}\ergscm$ per pixel ($\Delta v =
69\kms$), and detection limits for (close to) unresolved features in
several hundred or more stacked spectra can easily reach an impressive
$10^{-18}\ergscm$.  In Figure 3 we have reproduced the co-added SDSS
spectra in the wavelength region of the \oii\ doublet for 345 \caii\
and 3461 strong \mgii\ absorption line galaxies with $0.4 < \zabs <
1.3$ showing clear detection of the nebular emission lines from \hii\
regions associated with the absorbers. Each spectrum was converted
into units of luminosity before combining, and the best two-component
Gaussian fits are overplotted. Composite spectra are shown for low
($<$$z$$>\simeq$0.6) and high ($<$$z$$>\simeq$1.0) redshift
sub-samples.

The emission line detections allow for the first time a direct
estimate of the average SFR in an absorption-selected galaxy
population. Using the conversion between \oii\ line luminosity and SFR
proposed by Kewley et al.\,\cite{2004AJ....127.2002K} results in
values of 0.11 (0.48)$\Msolyr$ for the low- (high-) redshift \caii\
samples (after correcting for dust attenuation) and 0.11
(0.14)$\Msolyr$ for the DLAs selected by strong Mg\,{\sc ii}
absorption.

These values apply to the observed SFR within the finite aperture of
the SDSS fibre, which is centered on the quasar and not on the
absorber host galaxy.  The proper transverse radius covered by a
3\,arcsec fibre is $\sim7.5\,h^{-1}{\rm kpc}$ at $z=0.6$ and
$\sim8.4\,h^{-1}{\rm kpc}$ at $z=1$, i.e.  comparable to the expected
size of the star forming extent of $L^*$ galaxies\,\footnote{A flat
cosmology with $\Omega_\Lambda=0.7$, $\Omega_M=0.3$,
$H_0=100\,h\,\kms\,{\rm Mpc}^{-1}$ is assumed throughout}.  Thus, the
fibre may be covering a fraction of empty sky, which has implications
for measures of SFR per unit area of the absorbers; or the fibre may
exclude a fraction of the galaxy light, with implications for
estimating the total SFR per absorber.  These results allow us to
place lower limits on the SFR per unit area of the absorbers; however,
further interpretation requires an estimate of the physical extent of
the absorbers.  Both observational\,\cite{2005MNRAS.364.1467Z} and
recent theoretical work predict mean impact parameters between low- to
intermediate-redshift DLAs and host galaxies of order $10h^{-1}\,$kpc,
i.e.  DLAs are caused by extended gaseous halos/disks around central
galaxies.  Adopting this model we estimate aperture corrections and
derive SFRs per unit area of $11$ ($36$)$\times
10^{-4} h^2\Msolyr\perkpcsq$ for the \caii\ absorbers and $14$
($11$)$\times 10^{-4} h^2\Msolyr\perkpcsq$ for \mgii-selected DLAs.

The SFRs per unit area for DLAs lie at least a factor of five below the
prediction of the Schmidt law\,\cite{1959ApJ...129..243S}, which
relates the surface density of neutral gas and SFR in galaxies.  A
straightforward calculation also shows that the contribution of both
the \mgii-selected DLAs and \caii\ absorbers to the global SFR density
of the universe at $z\lsim 1$ is $<10$\%.  Our results are
dramatically different from those of Hopkins
et~al.~(2005)\,\cite{2005ApJ...630..108H}, who proposed that DLA absorbers
are responsible for $\sim 80\%$ of the global SFR density, $\sfrd$, at
redshifts $z \lsim 1$.  Hopkins et~al.  assumed that the Schmidt law
holds for DLA absorbers and the difference with our results is entirely
attributable to the much lower, directly determined, SFR per unit area
of the DLAs.

Interpretation of the results for the \caii\ absorbers is less certain because
their mean \nhi \ column densities are not yet established and the results of 
imaging studies to determine impact parameters and luminosities of their host
galaxies are awaited. 

An attractive explanation for the low observed SFR per unit area in the DLAs is
that the threshold \hi\ column density required to trigger star formation is
significantly higher than in the nearby galaxies where the empirical Schmidt law
was calibrated.  Whatever the physical explanation, it is now difficult to
escape the conclusion that only a small fraction of the star formation rate seen
directly in galaxy surveys at redshifts $z \sim 1$ is associated with DLAs.
Evidently, the largest contribution to the DLA cross-section is from gas which
is too diffuse to support high rates of star formation and metal production,
thus explaining the generally low metallicities of most DLAs.

\section*{Acknowledgments}
VW is supported by the MAGPOP EU Marie Curie Training and Research Network.

\section*{References}
%\bibliography{../paper_sfr/refs_all}
%\bibliography{v_wild_v2}
%\begin{thebibliography}{99}

%\end{thebibliography}
%\include{bibfile}

\end{document}